\documentclass{iau}
\usepackage{graphicx}

\title[IAUS~304.~Relative frequencies of SNe versus properties of spiral hosts] 
{Relative frequencies of supernovae\\ versus properties of spiral hosts}

\author[A.~A.~Hakobyan et al.] 
{A.~A.~Hakobyan$^1$,
T.~A.~Nazaryan$^1$,
V.~Zh.~Adibekyan$^2$,
A.~R.~Petrosian$^1$,
L.~S.~Aramyan$^1$,
D.~Kunth$^3$,
G.~A.~Mamon$^3$,
V.~de~Lapparent$^3$,
E.~Bertin$^3$,
J.~M.~Gomes$^2$,
\and M.~Turatto$^4$}

\affiliation{$^1$Byurakan Astrophysical Observatory, 0213 Byurakan, Aragatsotn Province, Armenia\\
email: {\tt hakobyan@bao.sci.am} \\[\affilskip]
$^2$Centro de Astrof\'{i}sica da Universidade do Porto, Rua das Estrelas, 4150-762 Porto, Portugal\\[\affilskip]
$^3$Institut d'Astrophysique de Paris, 98 bis Bd Arago, 75014 Paris, France\\[\affilskip]
$^4$INAF-Osservatorio Astronomico di Padova, Vicolo dell'Osservatorio 5, 35122 Padova, Italy}

\pubyear{2013}
\volume{304}  
\pagerange{339--340}
\jname{Multiwavelength AGN Surveys and Studies}
\editors{A.~M.~Mickaelian \& D.~B.~Sanders, eds.}
\begin{document}

\maketitle

\begin{abstract}
  In this work, we present an analysis of SNe number ratios in spiral galaxies
  with different morphological subtypes, luminosities, sSFR, and metallicities,
  to provide important information about the physical properties of
  the progenitor populations.
\end{abstract}


We investigate the morphological dependence of the number ratios of various SN types,
using a large sample of SNe along with information about magnitudes,
sSFR, and metallicities of their spiral host galaxies.
Our sample of 692 nearby SNe ($\leq97~{\rm Mpc}$) within 608 host galaxies is
selected from the database of SNe and their host galaxies
presented in \cite{2012A&A...544A..81H}.

\begin{figure}
  \begin{center}$
  \begin{array}{cc}
  \includegraphics[width=0.47\hsize]{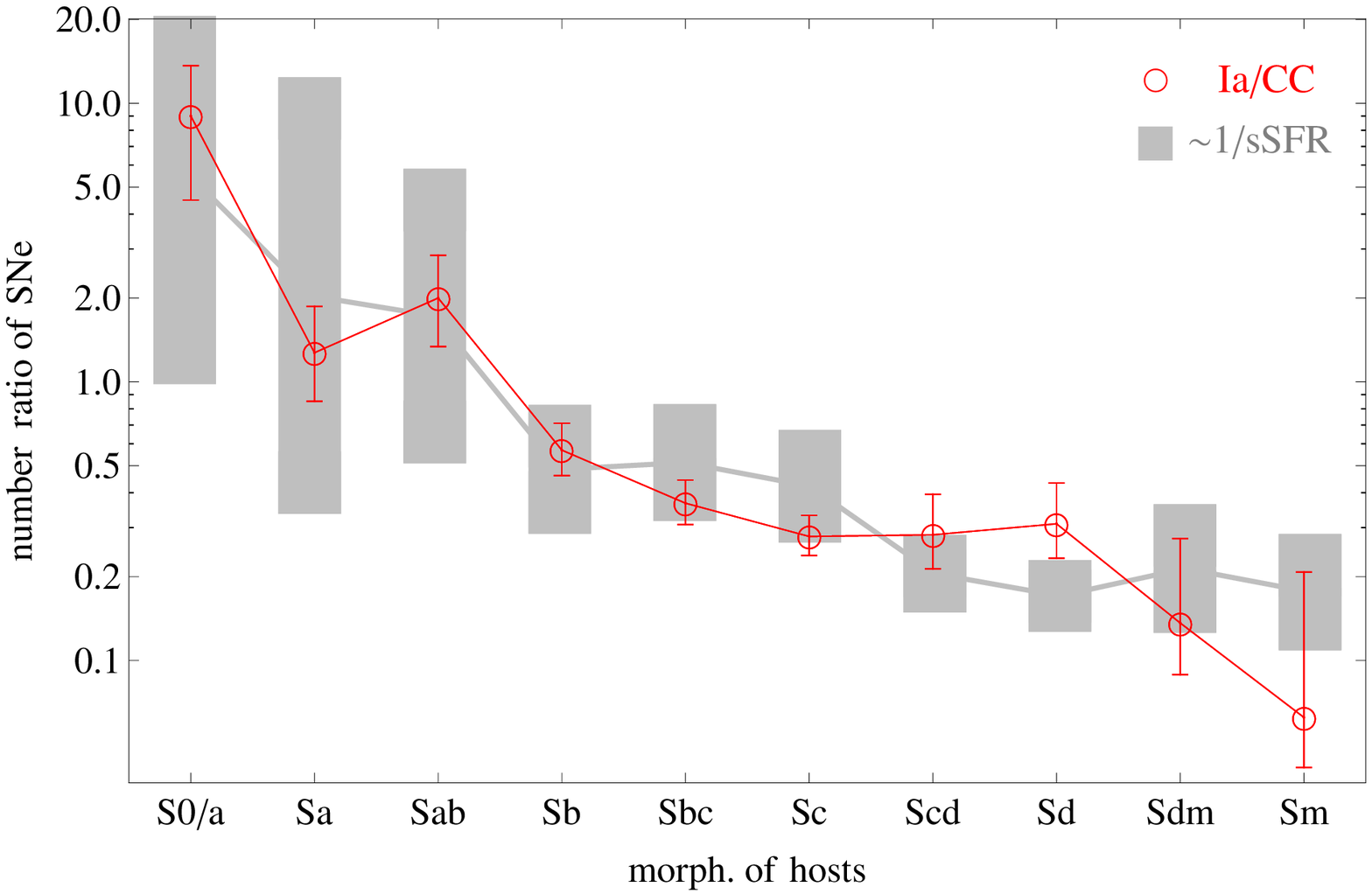} &
  \includegraphics[width=0.47\hsize]{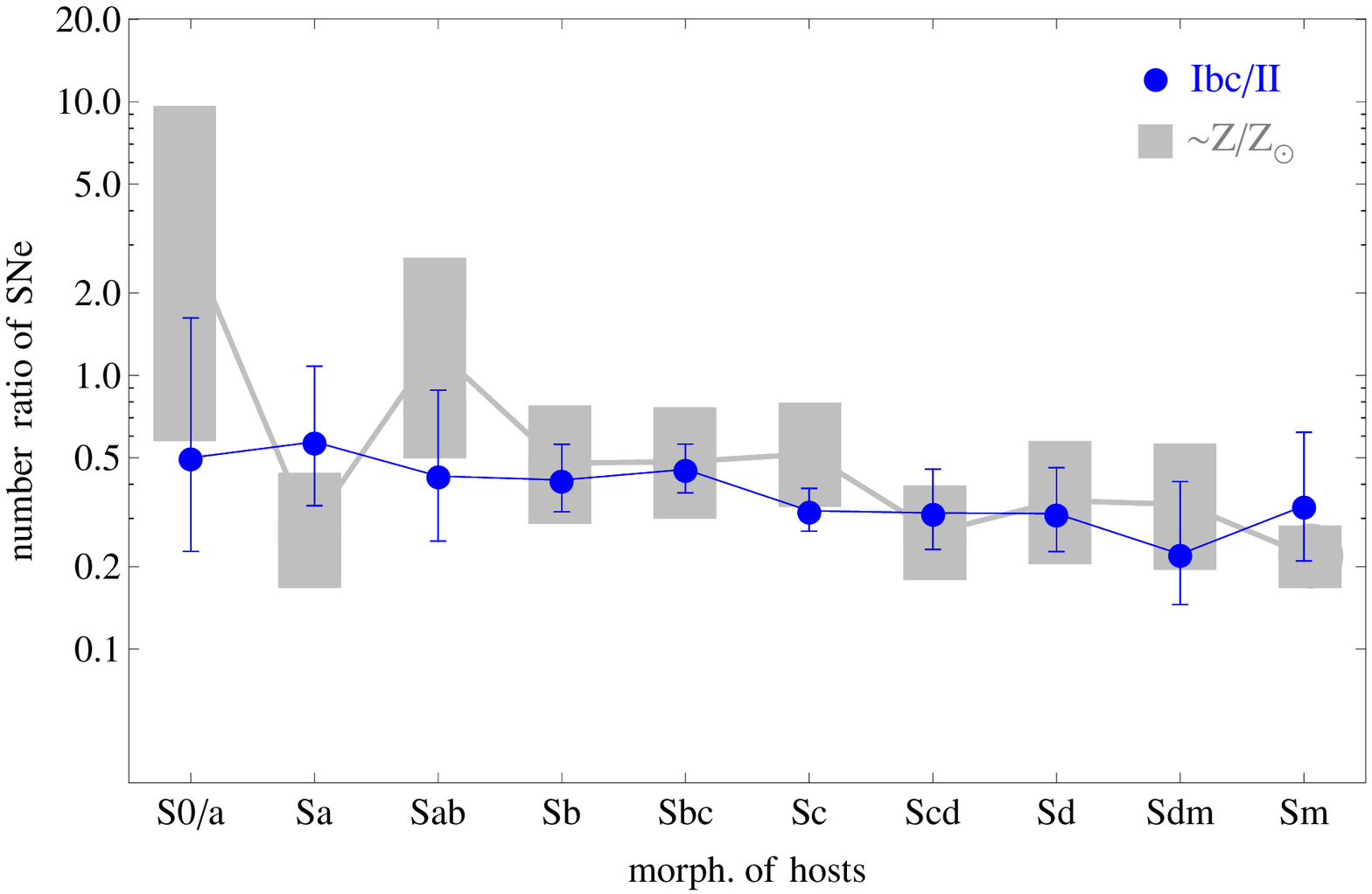}
  \end{array}$
  \end{center}
  \caption{Relative frequency of SNe types as a function of
  host morphology. The $N_{\rm Ia}/N_{\rm CC}$ is presented by red open circles (left), while the
  $N_{\rm Ibc}/N_{\rm II}$ is presented by blue filled circles (right).
  The distributions of 1/sSFR (left) and ${\rm Z/Z_\odot}$ (right) are presented
  by grey rectangles and shifted towards the vertical axis to visually
  fit the $N_{\rm Ia}/N_{\rm CC}$ and $N_{\rm Ibc}/N_{\rm II}$ distributions,
  respectively.}
  \label{ratiottype}
\end{figure}

As can be seen from the left panel of Fig.~\ref{ratiottype},
there is a strong trend in behavior of $N_{\rm Ia}/N_{\rm CC}$
depending from host galaxy morphological types,
such that early-type spirals
include proportionally more Ia SNe.
To demonstrate the relation between $N_{\rm Ia}/N_{\rm CC}$ and sSFR,
in the left panel of Fig.~\ref{ratiottype}
we have also shown the distribution of 1/sSFR according
to the morphologies of the host galaxies.
The best-fit sSFR values are extracted
for host galaxies of 253 SNe available from the SDSS.
Again, as for $N_{\rm Ia}/N_{\rm CC}$, there is strong trend for
the distribution of 1/sSFR, such that sSFR of host galaxies
systematically increased from early- (high-mass/luminosity) to
late-type (low-mass/luminosity) spirals.
Here, we share the view with \cite{2009A&A...503..137B}
that massive (early-type) spirals have, on average, lower sSFR.
Therefore, the behavior of $N_{\rm Ia}/N_{\rm CC}$ versus morphology
is simply reflection of the behavior of 1/sSFR versus morphological types of galaxies.

The right panel of Fig.~\ref{ratiottype} presents the distribution of
$N_{\rm Ibc}/N_{\rm II}$ versus host morphology.
The distribution is mostly flat and shows no dependence from the morphological types.
However, when we divide the host sample into two broad morphology bins,
i.e., S0/a--Sbc and Sc--Sm,
the difference between number ratios in these bins becomes barely
significant.
The trend is same for $N_{\rm Ic}/N_{\rm Ib}$, such that early-type
spirals include proportionally more type Ic than type Ib SNe.
Thus, we have seen that $N_{\rm Ibc}/N_{\rm II}$ and $N_{\rm Ic}/N_{\rm Ib}$
ratios show weak variation with morphology of hosts.

In general, the $N_{\rm Ibc}/N_{\rm II}$ depends on metallicity, age, and
fraction of binary systems (\cite[Eldridge et al. 2008]{2008MNRAS.384.1109E}).
To qualitatively present relation between this ratio and metallicity,
we extracted the best-fit metallicities available
only for 196 CC SNe host galaxies from the SDSS.
In the right panel of Fig.~\ref{ratiottype},
we presented metallicity values in the fixed morphology bins by grey
rectangles.
Here, we have seen that metallicity show no
or low variation with morphology of spiral hosts.
Therefore, the behaviors of $N_{\rm Ibc}/N_{\rm II}$ and $N_{\rm Ic}/N_{\rm Ib}$
ratios versus morphology can be interpreted by simple reflection of
the behavior of metallicity versus morphological types of spirals.

From our results, it is also clear that
the $N_{\rm Ia}/N_{\rm CC}$ is higher for brighter galaxies.
The result is in agreement with that of \cite{2009A&A...503..137B}:
the brighter, i.e., high mass, galaxies host proportionally
more Ia than CC SNe.
In both cases, the results are not highly significant.
The $N_{\rm Ibc}/N_{\rm II}$ and $N_{\rm Ic}/N_{\rm Ib}$ ratios are
higher in brighter galaxies. These
results are similar to $N_{\rm Ibc}/N_{\rm II}$ versus $M_{\rm B}$ relations
obtained in \cite{2009A&A...503..137B}.
Here, we agree with their interpretation:
with increasing of luminosity (metallicity), the stellar envelope is
more easily lost and lower mass stars may become type Ibc (Ic) SNe,
increasing thus the $N_{\rm Ibc}/N_{\rm II}$ ($N_{\rm Ic}/N_{\rm Ib}$) ratio.

It is also evident that despite the degree of subjectivity
involved in the morphological classifications,
the number ratio-morphology relation is probably tighter than the
number ratio-luminosity (metallicity) relation
(according to significance values).
The morphology that we have done in \cite{2012A&A...544A..81H}
using RGB images of the SDSS, i.e., considering to some extent also
the color of galaxies, is more directly related to stellar population
than the indirectly estimated metallicity.
In particular, 64\% of type Ia SNe are located in S0/a--Sbc
galaxies in contrast to 40\% of CC SNe.
The mean morphological subtype of spiral hosts of Ia SNe is significantly earlier than
those of all types of CC SNe hosts, and can be interpreted as the main factor constraining
sSFR and metallicity of the progenitor populations.
The complete study will be presented in \cite{hakobyan2014}.

\begin{acknowledgement}
A.~R.~P., A.~A.~H., and L.~S.~A. are supported by the Research Project
of the SCS and CNRS.
This work was made possible in part by a research grant from the
Armenian National Science and Education Fund based in New York, USA.
This work was supported by State Committee Science MES RA,
in frame of the research project number 13-1C013.
V.~Zh.~A. is supported by grant SFRH/BPD/70574/2010 from FCT (Portugal).
\end{acknowledgement}


\begin{thebibliography}{}

\bibitem[Boissier \& Prantzos (2009)]{2009A&A...503..137B}
{Boissier, S., \& Prantzos, N.} 2009, \textit{A\&A}, 503, 137

\bibitem[Eldridge et al.(2008)]{2008MNRAS.384.1109E}
{Eldridge,~J.~J., Izzard,~R.~G., \& Tout,~C.~A.} 2008, \textit{MNRAS}, 384, 1109

\bibitem[Hakobyan et al. (2012)]{2012A&A...544A..81H}
{Hakobyan,~A.~A., Adibekyan,~V.~Z., Aramyan,~L.~S., et al.} 2012, \textit{A\&A}, 544, A81

\bibitem[Hakobyan et al. (2014)]{hakobyan2014}
{Hakobyan,~A.~A., Nazaryan,~T.~A., Adibekyan,~V.~Z., et al.} 2014, \textit{MNRAS}, 444, 2428

\end{thebibliography}
\end{document}